
\magnification=1200
\hsize=6truein\vsize=8.5truein
\font\open=msbm10 
\font\goth=eufm10  

\font\bigbf=cmbx10 scaled\magstep1

\font\ssb=cmssbx10
\def\mbox#1{{\leavevmode\hbox{#1}}}
\def\oR{\mbox{\open\char82}}

\def\gD{\mbox{{\goth\char68}}}
\def\ssC{\mbox{{\ssb\char67}}}
\def\ssD{\mbox{{\ssb\char68}}}

\def\S{{\rm S}}

\def\la{\lambda}

\def\det{{\rm det\,}}
\def\Real{{\rm Re\,}}
\def\zf{$\zeta$--function}
\def\zfs{$\zeta$--functions}

\def\frac#1/#2{\leavevmode\kern.1em
\raise.5ex\hbox{\the\scriptfont0 #1}\kern-.1em/\kern-.15em
\lower.25ex\hbox{\the\scriptfont0 #2}}
\def\sfrac#1/#2{\leavevmode\kern.1em
\raise.5ex\hbox{\the\scriptscriptfont0 #1}\kern-.1em/\kern-.15em
\lower.25ex\hbox{\the\scriptscriptfont0 #2}}

\def\gtorder{\mathrel{\raise.3ex\hbox{$>$}\mkern-14mu
             \lower0.6ex\hbox{$\sim$}}}
\def\ltorder{\mathrel{\raise.3ex\hbox{$<$}|mkern-14mu
             \lower0.6ex\hbox{\sim$}}}

\def\semidirprod{\rlap{\ss C}\raise1pt\hbox{$\mkern.75mu\times$}}

\def\for{\lower6pt\hbox{$\Big|$}}
\def\fish{\kern-.25em{\phantom{abcde}\over \phantom{abcde}}\kern-.25em}

\def\boxit#1{\vbox{\hrule\hbox{\vrule\kern3pt
        \vbox{\kern3pt#1\kern3pt}\kern3pt\vrule}\hrule}}
\def\dalemb#1#2{{\vbox{\hrule height .#2pt
        \hbox{\vrule width.#2pt height#1pt \kern#1pt
                \vrule width.#2pt}
        \hrule height.#2pt}}}


\def\noin{\noindent}

\def\al{\alpha}
\def\be{\beta}
\def\ga{\gamma}

\def\Ga{\Gamma}

\def\ep{\epsilon}

\def\ka{\kappa}
\def\la{\lambda}

\def\th{\theta}
\def\ze{\zeta}
\def\De{\Delta}

\def\comb#1#2{{\left(#1\atop#2\right)}}

\def\eg{{\it e.g. }}
\def\ie{{\it i.e. }}
\def\cf{{\it cf }}
\def\pa{\partial}

\def\gap{\vskip 20truept}

\def\sumdash#1{{\mathop{{\sum}'}_{#1}}}
\def\ref{\smallskip\par\noindent}
\def\ttimes{\!\times\!}

\def\sect{{\vskip 10truept\noindent}}

\def\3j#1#2#3#4#5#6{\left\lgroup\matrix{#1&#2&#3\cr#4&#5&#6\cr}
\right\rgroup}


\def\nolabels{\def\eqnlabel##1{}\def\eqlabel##1{}\def\reflabel##1{}}
\def\writelabels{\def\eqnlabel##1{%
{\escapechar=` \hfill\rlap{\hskip.09in\string##1}}}%
\def\eqlabel##1{{\escapechar=` \rlap{\hskip.09in\string##1}}}%
\def\reflabel##1{\noexpand\llap{\string\string\string##1\hskip.31in}}}
\nolabels
\global\newcount\meqno \global\meqno=1
\global\meqno=1
\def\eqnn#1{\xdef #1{(\the\meqno)}%
\global\advance\meqno by1\eqnlabel#1}
\def\eqna#1{\xdef #1##1{\hbox{$(\the\meqno##1)$}}%
\global\advance\meqno by1\eqnlabel{#1$\{\}$}}
\def\eqn#1#2{\xdef #1{(\the\meqno)}\global\advance\meqno by1%
$$#2\eqno#1\eqlabel#1$$}


\global\newcount\refno
\global\refno=1 \newwrite\reffile
\newwrite\refmac
\newlinechar=`\^^J
\def\ref#1#2{\the\refno\nref#1{#2}}
\def\nref#1#2{\xdef#1{{\bf\the\refno}} 
\ifnum\refno=1\immediate\openout\reffile=refs.tmp\fi
\immediate\write\reffile{
     \noexpand\item{[{\noexpand#1}]\ }#2\noexpand\nobreak.}
     \immediate\write\refmac{\def\noexpand#1{\the\refno}}
   \global\advance\refno by1}
\def\semi{;\hfil\noexpand\break ^^J}
\def\refn#1#2{\nref#1{#2}}

\def\cmp#1#2#3{{\it Comm. Math. Phys.} {\bf {#1}} (19{#2}) #3}

\def\np#1#2#3{{\it Nucl. Phys.} {\bf B{#1}} (19{#2}) #3}

\def\prD#1#2#3{{\it Phys. Rev.} {\bf D{#1}} (19{#2}) #3}


\refn\Chang{Peter Chang and J.S.Dowker \np {395}{93}{407}}
\refn\Barnes{E.W.Barnes {\it Trans.Camb.Phil.Soc.} (1903) 376}
\refn\Allen{B.Allen, PhD Thesis, University of Cambridge, 1984}
\refn\Chodos{A.Chodos and E.Myers {\it Can.J.Phys.} {\bf 64} (1986) 633}
\refn\Bateman{A.Erdelyi,W.Magnus,F.Oberhettinger and F.G.Tricomi {\it Higher
Transcendental functions} McGraw-Hill, New York (1953)}
\refn\Berard{P.B\'erard \& G.Besson {\it Ann.Inst.Fourier}
{\bf 30}, (1980) 237}
\refn\Coxetera{H.S.M.Coxeter {\it Regular Polytopes} Methuen, London (1948)}
\refn\Coxeterb{H.S.M.Coxeter {\it Regular Complex Polytopes} 2nd. Edn.
Cambridge University Press, Cambridge (1991)}
\refn\Polya{G.Polya and B.Meyer {\it Comptes Rend. Acad.Sci. (Paris)}
{\bf 228} (1949) 28}
\refn\Meyer{B.Meyer {\it Can.J.Math.} {\bf 6} (1954) 135}
\refn\Coxeterc{H.S.M.Coxeter and W.O.J.Moser {\it Generators and
relations for finite groups} Springer-Verlag, Berlin (1957)}
\refn\Weisbergerb{W.I.Weisberger \cmp{112}{87}{633}}
\refn\Hortacsu{M.Hortacsu, K.D.Rothe and B.Schroer \prD {20}{79}{3203}}
\refn\Weisbergera{W.I.Weisberger \np{284}{87}{171}}
\refn\Cesaro{G.Cesaro {\it Mineralogical Mag.} {\bf 17} (1915) 173}
\refn\Fedorov{E.S.Fedorov {\it Mineralogical Mag.} {\bf 18} (1919) 99}
\refn\Fedosov{B.V.Fedosov {\it Sov.Mat.Dokl.} {\bf 4} (1963) 1092; {\it ibid}
{\bf 5} (1964) 988}
\refn\Coxeterd{H.S.M.Coxeter and G.J.Whitrow {\it Proc. Roy. Soc.} {\bf A200}
(1950) 417}


\vglue 1truein
\rightline {MUTP/93/15}
\gap
\centerline {\bigbf Effective action in spherical domains}
\vskip 15truept
\centerline{J.S.Dowker}
\vskip 10 truept
\centerline {Department of Theoretical Physics,}
\centerline{The University of Manchester, Manchester, England.}
\vskip 40truept
\centerline {Abstract}
\vskip 10truept
The effective action on an orbifolded sphere is computed for
minimally coupled scalar fields. The results are presented in terms of
derivatives of Barnes \zfs\ and it is shown how these may be evaluated.
Numerical values are shown. An analytical, heat-kernel derivation of
the Ces\`aro-Fedorov formula for the number of symmetry planes of a
regular solid is also presented.
\vfill\eject
\noin{\bf 1. Introduction}.

\noin In earlier work [\Chang] we have shown that the \zf, $\ze_\Ga(s)$, on
orbifold-factored
spheres, S$^d/\Ga$, for a conformally coupled scalar field, is given by a
Barnes \zf, [\Barnes], $\zeta_d(s,a\mid{\bf {d}})$, where the $ d_i$ are the
{\it
degrees} associated with the tiling group $\Ga$. The free-field Casimir energy
on the
space-time $\oR\times$S$^d/\Ga$ was given as the value of the \zf\ at a
negativ
 e
integer which evaluated to a generalised Bernoulli function. In the present
work we wish to consider the effective action on orbifolds S$^d/\Ga$
which this time are to be looked upon as Euclidean space-times. In
particular we will discuss $d=2$ and $d=3$, concentrating on the former.

The simplifying assumption in our previous work was that of conformal coupling
on $\oR\times$S$^d/\Ga$. This made the relevant eigenvalues perfect squares
and allowed us to use known generating functions  to incorporate the
degeneracies. From the point of view of field theories on the space-times
S$^d/\Ga$, retaining this assumption would be rather artificial.
A more appropriate choice would be minimal coupling, or
possibly conformal coupling, on S$^d/\Ga$. (These coincide for $d=2$.)

The quantities in which we are interested are $\ze_\Ga'(0)$ and
$\ze_\Ga(0)$. The latter determines the divergence in the effective
action and the former is, up to a factor and a finite addition, the
renormalised
effective action (\ie half the logarithm of the functional determinant).
\sect{\bf 2. Eigenvalues, degeneracies and zeta functions}.

\noin For the aforementioned conformal coupling, the eigenvalues of the second
order operator $-\De_2+\xi R\,\,\, (\xi=(d-1)/4d)$ are
\eqn\eigena{\la_n={1\over4}(n+d-2)^2}
with degeneracies that we shall leave unspecified here.

In our previous work [\Chang] we showed that the corresponding Neumann
and Dirichlet
\zfs\ on S$^d/\Ga$ were,
\eqn\nzf{
\zeta^{(C)}_{{}_N}(s)=\zeta_d\left(2s,(d-1)/2\mid {\bf d}\right),}
\eqn\dzf{
\zeta^{(C)}_{{}_D}(s)=\zeta_d\left(2s,{\textstyle\sum}\,d_i-(d-1)/2\mid {\bf
d}
\right),}
where the general definition of the Barnes \zf\ is
\eqn\bzeta{\eqalign{
\zeta_d(s,a\mid{\bf {d}})&={i\Gamma(1-s)\over2\pi}\int_L d\tau {\exp(-a\tau)
(-\tau)^{s-1}\over\prod_{i=1}^d\big(1-\exp(-d_i\tau)\big)}\cr
&=\sum_{{\bf {m}}={\bf 0}}^\infty{1\over(a+{\bf {m.d}})^s},\qquad
\Real\, s>d.\cr}}

This shows that the eigenvalues are given specifically by
\eqn\eigenb{
\la_n=(a+{\bf m.d})^2} the degeneracies coming from coincidences.
The parameter $a$ is $(d-1)/2$ in the Neumann case and comparison with the
previous form shows that the integer $n= 2{\bf m.d}+1$, ${\bf m=0}$ upwards.
For Dirichlet conditions, $a=\sum d_i-(d-1)/2$ and then
$n=2{\bf m.d}-1$ with ${\bf m}=(1,1)$
upwards. The interpretation in two dimensions is that
the angular momentum is $L={\bf m.d}$ for Neumann and
${\bf m.d}-1$ for Dirichlet conditions.

Turning to minimal coupling, ($\xi=0$),
the eigenvalues of the Laplacian are
\eqn\mineigen{
\la_n=(a+{\bf m.d})^2-{(d-1)^2\over4}.}
and the corresponding \zf\ is
\eqn\minnzeta{
\ze(s)=\sum_{\bf m} {1\over\big((a+{\bf m.d})^2-(d-1)^2/4\big)^s}.
}
The origin ${\bf m}={\bf 0}$ is to be omitted for Neumann conditions, when
the \zf\ is denoted by $\bar\ze(s)$.

Consider a sum of the form
\eqn\genzet{
\ze(s)=\sum_{\bf m} {1\over\big((a+{\bf m.d})^2-\al^2\big)^s}
}
so that
\eqn\reln{
\bar\ze(s)=\ze(s)-(a^2-\al^2)^{-s}.
}
For minimal coupling, $\al=(d-1)/2$, while for conformal coupling in
$d$--dimensions, $\al=1/2$. We concentrate on minimal coupling.

A standard way of obtaining information about an expression such as
\genzet\ is to perform
a binomial expansion to produce a sum of known \zfs, in the present case a
sum of Barnes \zfs,
\eqn\binsum{
\ze(s)=\sum_{r=0}^\infty\al^{2r}{s(s+1)\ldots(s+r-1)\over
r!}\ze_d(2s+2r,a\mid
{\bf d}).
}
{}From this, the value of $\ze(s)$ at a nonpositive integer is easily found.
For example the important value $\ze(0)$ is given by
\eqn\zetazero{
\ze(0)=\ze_d(0,a\mid{\bf d})+{1\over2}\sum_{r=1}^u{\al^{2r}\over r}N_{2r}(d)
}
and, more generally, we have
\eqn\zetaminusn{
\ze(-n)=\sum_{r=0}^n (-\al^2)^r\left(n\atop r\right)\ze_d(2r-2n,a\mid{\bf
d})+{(-1)^n\over2}\sum_{r=1}^u{n!(r-1)!\over(r+n)!}\al^{2n+2r}N_{2r},
}
where $u=d/2$ if $d$ is even and $u=(d-1)/2$ if $d$ is odd.

$N_r(d)$ is the residue defined by
\eqn\zelim{
\ze_d(s+r,a\mid{\bf d})\rightarrow {N_r(d)\over s}+R_r(d)
\quad{\rm as}\,\,s\rightarrow 0,
}
where $1\le r\le d$. Expressions for the residue and remainder involve
generalised Bernoulli functions and can be found in Barnes [\Barnes]. For
shortness, their dependence on the parameter $a$ is not indicated.

The form of the residues given by Barnes [\Barnes] is
$$N_r(d)={(-1)^{r+d}\over(r-1)!}\!\!\phantom{H}_dS_1^{(r+1)}(a)$$
where$\phantom{H}_dS_1^{(r+1)}(a)$ is the $(r+1)$-th derivative of
Barnes' generalised Bernoulli polynomial $\!\!\!\phantom{H}_dS_1(a)$.
The general relation with the more usual polynomials, [\Bateman], will
not be given here. Specific forms are
$$\phantom{H}_dS_1^{(d+1)}(a)={1\over\prod d_i},\quad
\phantom{H}_dS_1^{(d)}(a)={2a-\sum d_i\over2\prod d_i},$$
\eqn\poleres{
\phantom{H}_dS_1^{(d-1)}(a)={1\over12\prod d_i}\big(6a^2-6a\sum d_i+
\sum d_i^2+3\sum_{i<j}d_id_j\big).
}
Barnes also gives the values
\eqn\barzero{
\ze_d(-n,a\mid{\bf d})={(-1)^d\over
n+1}\!\!\!\!\phantom{H}_dS_{1+n}^{(1)}(a)=
{(-1)^d\over\prod d_i}{n!\over(d+n)!}B^{(d)}_{d+n}(a\mid{\bf d}).
}
{}From \zetazero, \poleres\ and \barzero\ we find, for two dimensions,
\eqn\confanom{
\ze_{{}_N}(0)=\ze_{{}_D}(0)={1\over12d_1d_2}\big(3-3(d_1+d_2)+(d_1+d_2)^2
+d_1d_2\big).
}

This corrects our previous expression [\Chang] obtained by an incorrect
manipulation of the heat-kernel.
\sect{\bf 3. The derivative of the zeta function}.

\noin The derivative at $s=0$ is a little more difficult to find. From
\binsum\ a first step is
\eqn\deri{
\ze'(0)=2\ze_d'(0,a\mid{\bf d})+\sum_{r=1}^u{\al^{2r}\over r}\left(R_{2r}+
{1\over2}N_{2r}\sum_1^{r-1}{1\over k}\right)
+\sum_{r=u+1}^\infty{\al^{2r}\over r}\ze_d(2r,a\mid{\bf d}).
}

The integral representation of the Barnes \zf\ allows the final sum in
\deri\ to be written as
$$\sum_{r=u+1}{\al^{2r}\over r\Ga(2r)}\int_0^\infty d\tau{\tau^{2r-1}
\exp(-a\tau)\over\prod_i\big(1-\exp(-d_i\tau)\big)}=$$
\eqn\derisum{
2\int_0^\infty\exp(-a\tau)\left(\cosh\al\tau-\sum_{r=0}^u{(\al\tau)^{2r}
\over(2r)!}\right){d\tau\over\tau\prod_i\big(1-\exp(-d_i\tau)\big)}.
}
In the Neumann case $a=\al$ and there is an infra-red, logarithmic
divergence at infinity caused by the zero mode which will be taken care of by
the transition to $\bar\ze$, \reln.

Although the integral converges nicely at $\tau=0$, the individual terms of
the integrand do not. It is enough to introduce another ultra-violet
analytic regularisation and define the intermediate quantity,
\eqn\uvrega{
2\int_0^\infty\exp(-a\tau)\left(\cosh\al\tau-\sum_{r=0}^u{(\al\tau)^{2r}
\over(2r)!}\right){\tau^{s-1}d\tau\over\prod_i\big(1-\exp(-d_i\tau)\big)}
}
whose $s=0$ limit gives \derisum.

After continuation, \uvrega\ integrates to
\eqn\uvregb{
\Ga(s)\big(\ze_d(s,a-\al\mid{\bf d})+\ze_d(s,a+\al\mid{\bf d})\big)-
2\sum_{r=0}^u {\al^{2r}\over(2r)!}\Ga(s+2r)\ze_d(s+2r,a\mid{\bf d}).
}

As $s$ tends to zero, each term in \uvregb\ yields a pole and a finite
remainder. The poles must cancel and so
\eqn\pole{
\ze_d(0,a-\al\mid{\bf d})+\ze_d(0,a+\al\mid{\bf d})-2\ze_d(0,a\mid{\bf d})
=\sum_{r=1}^u{\al^{2r}\over r}N_{2r}(d).
}
This condition is an identity between generalised Bernoulli functions.
Combinined with \zetazero\ it produces the symmetrical expression
\eqn\zetazerob{
\ze(0)={1\over2}\big(\ze_d(0,a-\al\mid{\bf d})+\ze_d(0,a+\al\mid{\bf d})\big).
}

The finite remainder in \uvregb\ is
$$\ze_d'(0,a-\al\mid{\bf d})+\ze_d'(0,a+\al\mid{\bf d})-2\ze_d'(0,a\mid
{\bf d})-\sum_{r=1}^u {\al^{2r}\over r}R_{2r}(d)-$$

\eqn\remaina{
\ga\big(\ze_d(0,a-\al\mid{\bf d})+\ze_d(0,a+\al\mid{\bf d})-
2\ze_d(0,a\mid{\bf d})\big)-
\sum_{r=1}^u {\al^{2r}\over r}\psi(2r)N_{2r}(d)
}
which, in view of \pole, can be written
$$\ze_d'(0,a-\al\mid{\bf d})+\ze_d'(0,a+\al\mid{\bf d})-2\ze_d'(0,a\mid
{\bf d})-$$
\eqn\remainb{
\sum_{r=1}^u {\al^{2r}\over r}R_{2r}(d)-\sum_{r=1}^u {\al^{2r}\over r}
\big(\psi(2r)+\ga\big)N_{2r}(d).
}
Combining this with \deri\ we have finally
\eqn\derib{
\ze'(0)=
\ze_d'(0,a-\al\mid{\bf d})+\ze_d'(0,a+\al\mid{\bf d})-
\sum_{r=1}^u {\al^{2r}\over 2r}\big(2\psi(2r)-\psi(r-1)+\ga\big)N_{2r}(d).
}
The fact that the remainders have cancelled, suggests that there is a more
rapid
route to this result.

Apart from the final term, \derib\ is the expression that
would have been obtained by a naive application of the `surrogate' \zf\ method
which is based on the product nature of the eigenvalues,
${(a-\al+\bf m.d})\,(a+\al+{\bf m.d})$, in \genzet\ followed by an application
of the rule
$\,\ln\det(AB)=\ln\det A+\ln\det B$. This method is suspect, as discussed by
Allen [\Allen] and by Chodos and Myers [\Chodos]. Allen [\Allen] derives a
particular `correction' term as in \derib. He also points out that
\zetazerob\ could be expected on the basis of the eigenvalue factorisation,
being the average of the regularised dimensions of the operator factors.

The final term in \derib\ can be rewritten

$$\sum_{r=1}^u {\al^{2r}\over2 r}\big(2\psi(2r)-\psi(r-1)+\ga\big)N_{2r}(d)=
\sum_{r=1}^u {\al^{2r}\over r}N_{2r}(d)
\sum_0^{r-1}{1\over 2k+1}.$$

In order to evaluate the effective action we must substitute the appropriate
values of $a$ and $\al$ for Neumann and Dirichlet conditions into \derib. In
the former case it is also necessary to remove the zero mode \ie to use
$\bar\ze$.
The relevant quantity then is the $\Ga$-modular form $\rho$, defined by,
[\Barnes],
\eqn\modform{
\lim_{\ep\rightarrow0}\ze_r'(0,\ep\mid{\bf d})=-\ln\ep-\ln\rho_r({\bf d}).
}

We find the following basic expressions
\eqn\zetaneu{
\ze_{{}_N}'(0)=-\ln\rho_d({\bf d})+\ze_d'(0,d-1\mid{\bf d})-\ln(d-1)-
\sum_{r=1}^u {\al^{2r}\over r}N_{2r}(d)\sum_0^{r-1}{1\over 2k+1}
}
and
\eqn\zetadir{
\ze_{{}_D}'(0)=\ze_d'(0,d_0\mid{\bf d})+\ze_d'(0,d_0+d-1\mid{\bf d})-
\sum_{r=1}^u {\al^{2r}\over2 r}N_{2r}(d)\sum_0^{r-1}{1\over 2k+1},
}
where $d_0=\sum_id_i-d+1$ is the number of reflecting planes in $\Ga$. We
recall that $\al=(d-1)/2$ for minimal coupling and that in \zetaneu,
$N$ is evaluated at $a=(d-1)/2$ and in \zetadir\ at $a=d_0+(d-1)/2$.

In the case of the two-sphere, \zetaneu\ and \zetadir\ become

\eqn\zetaneutwo{
\ze_{{}_N}'(0)=-\ln\rho_2({\bf d})+\ze_2'(0,1\mid{\bf d})-{1\over4g}
}
and
\eqn\zetadirtwo{
\ze_{{}_D}'(0)=\ze_2'(0,d_0\mid{\bf d})+\ze_2'(0,d_0+1\mid{\bf d})-{1\over4g}
}
where we have set $g=d_1d_2$, the order of the rotational part of $\Ga$.

For the three-sphere

\eqn\zetaneuthree{
\ze_{{}_N}'(0)=-\ln\rho_3({\bf d})+\ze_3'(0,2\mid{\bf d})-\ln2+{d_0\over2g}
}
and
\eqn\zetadirthree{
\ze_{{}_D}'(0)=\ze_3'(0,d_0\mid{\bf d})+\ze_3'(0,d_0+2\mid{\bf d})-{d_0\over2g}
}
where $g=d_1d_2d_3$. We note the change of sign in the last term.

Equations \zetaneutwo\ to \zetadirthree\ are the calculational formulae we
shall use in the rest of this paper. It is also possible to evaluate the
derivative of the \zf\ at negative integers, $\ze'(-n)$. This would be
relevant if we were interested in the effective action on product spaces like
$\oR\times\oR^k\times \S^d/\Ga$. A few details are presented in the appendix.

Although our main interest is in minimal coupling, it should be mentioned
that the result \derib\ can be used immediately for massive fields,
assuming that the appropriate value of $\al$ is real. This means that the mass
$\kappa$ is restricted to the region $0\le \kappa\le (d-1)/2$. For larger
masses a slightly different continuation is needed.

\sect {\bf 4. The derivative of the Barnes zeta function}.

\noin We turn now to the evaluation of the derivatives needed
in \zetaneutwo\ and \zetadirtwo. A preliminary step is to remove any common
factors of the degrees $d_1$ and $d_2$ by setting $d_i=ce_i$ with $e_1$ and
$e_2$ coprime so that the denominator function in \bzeta\ equals
$c(b+{\bf m.e})$ where $b=a/c$.

The summation in \bzeta\ is rewitten by introducing the residue classes with
respect to ${\bf e}$. On setting
$$m_1=n_1e_2+p_2,\quad m_2=n_2e_1+p_1,$$ where
$0\le p_i\le e_i-1$, the denominator function in \bzeta\ equals
$c\big(b+e_1e_2(n_1+n_2)+ p_2e_1+ p_1e_2\big)$
and the sum over ${\bf m}$ becomes
\eqn\zetaa
{\eqalign{\ze_2(s,a\mid{\bf d})&=c^{-s}\sum_{\bf p}\sum_{n=0}^\infty{1+n
\over (b+e_1e_2n+p_2e_1+ p_1e_2)^s}\cr
&={c^{2-s}\over g}\sum_{{\bf p},n}{1\over(b+N)^{s-1}}+\left({c\over
g}\right)!
 !s
\sum_{\bf p}(1-w_b)\sum_{n=0}^\infty{1\over(n+w_b)^s},\cr}
}
where
$$N=e_1e_2n+p_2e_1+p_1e_2,\quad{\rm and}\quad w_b={b\over e_1e_2}+
{p_1\over e_1}+{p_2\over e_2}.$$

Consider the integer $N= e_1e_2n+p_2e_1+ p_1e_2$. As $n$ ranges over $0$
to $\infty$, and the $p_i$ over their domains, $N$ will likewise run over
this infinite range with the exception of
some integers $< e_1e_2$ at the beginning, specifically those integers that
equal $p_2e_1+ p_1e_2$ mod$\,e_1e_2$ for $p_2e_1+ p_1e_2>e_1e_2$. We denote
these missing integers by $\nu_i$. Apart from these terms,  the first sum
in the second line of \zetaa\ will immediately give a single Hurwitz \zf,
\eqn\zetab{
\ze_2(s,a\mid{\bf d})={c^{2-s}\over
g}\left(\ze_R(s-1,b)-\sum_i{1\over(b+\nu_i)^{s-1}}\right)
+\left({c\over g}\right)^s\sum_{\bf p}(1-w_b)\ze_R(s,w_b).
}

\zetab\ is a convenient form for numerical evaluation. It provides an
explicit analytical continuation of this {\it integral} Barnes \zf.

For the derivative at $s=0$ we find, after inserting the known values of the
Hurwitz \zf\ and its derivative,

\eqn\dzetaa{\eqalign{
\ze_2'(0,a\mid{\bf d})&={c^2\over
g}\left(\ze_R'(-1,b)+\sum_i(\nu_i+b)\ln(\nu_i+b)\right)\cr
&-{c^2\over g}\left(\ze_R(-1,b)-\sum_i(b+\nu_i)\right)\ln c\cr
&+\sum_{\bf p}(1-w_b)\left(\ln\big(\Ga(w_b)/\sqrt(2\pi)\big)-(1/2-w_b)
\ln(g/c)\right).\cr}
}

Letting $a$ tend to zero in \dzetaa\ and comparing with the definition of
$\ln\rho$, \modform, one finds that
\eqn\modformb{\eqalign{
\ln\rho_2({\bf d})&=-{c^2\over g}\left(\ze_R'(-1)+\sum_i\nu_i\ln\nu_i\right)
-{c^2\over g}\left({1\over12}+\sum_i\nu_i\right)\ln c\cr
&-\sumdash{\bf p}(1-w_0)\left(\ln\big(\Ga(w_0)/\sqrt(2\pi)\big)-
({1\over2}-w_0)\ln (g/c)\right)-{1\over2}\ln(g/2\pi c)\cr
}
}
where $w_0=p_1/e_1+p_2/e_2$ and the dash means that the term $p_1=p_2=0$
is to be omitted from the sum.

Barnes gives a formula in terms of the multiple $\Ga$-function,
\eqn\multgam{
\ze_r'(0,a\mid{\bf d})=\ln\left({\Ga_r(a)\over\rho_r({\bf d})}\right).}

Formal expressions for the functional determinants are thus
\eqn\funcda{
e^{-\ze'_N(0)}=e^{1/4g^2}{\rho_2^2({\bf d})\over\Ga_2(1)}
}
and
\eqn\funcdb{
e^{-\ze'_D(0)}=e^{1/4g^2}{\rho_2^2({\bf d})\over\Ga_2(d_0)\Ga_2(d_0+1)}.
}
Our results, \zetab\ and \dzetaa, \modformb,  can be thought of as
computational

formulae for these functions in terms of simpler ones.

It is not necessary to rearrange the summation as in \zetaa.
We have done so in order to extract the term $\ze_R(s-1,b)$. If the summation
is left as in the first line of \zetaa, it can immediately be turned into
a sum of Hurwitz \zfs,
\eqn\altsum{
\ze_2(s,a\mid{\bf d})=\left({c\over g}\right)^s\sum_{\bf p}\big(
\ze_R(s-1,w_b)+(1-w_b)\ze_R(s,w_b)\big).
}
Then we have the alternative form
\eqn\altdiff{
\eqalign{
\ze'_2(0,a\mid{\bf d})
&=\ln(c/g)\sum_{\bf
p}\big(\ze_R(-1,w_b)+(1-w_b)\ze_R(0,w_b)\big)\cr
&+\sum_{\bf p}\big(\ze'_R(-1,w_b)+(1-w_b)\ze'_R(0,w_b)\big)\cr
&={1\over12g}\big(6a^2-6a(d_0+1)+(d_0+1)^2+g\big)\ln(c/g)\cr
&+\sum_{\bf p}\big(\ze'_R(-1,w_b)+(1-w_b)\ln\big(\Ga(w_b)
/\sqrt(2\pi)\big)\big).\cr
}
}
In this way we do not need to find the missing integers (nor even the common
factor $c$) but the price is
the multiple evaluation of $\ze'_R(-1,w_b)$ by a numerical procedure.
There is no difficulty in this but \zetab\ is faster and more accurate.
Equation

\altdiff\ constitutes a useful check.
\sect{\bf 5. The point groups}.

\noin A limited test of our formulae is provided by the dihedral case,
$\Ga=[q]$ in Coxeter's notation [\Coxetera,\Coxeterb]. (Sch\"onflies
would write $C_{qv}$ and it is $\ssC_q[\ssD_q$ in Polya and Meyer
[\Polya,\Meyer]. Table 2 in [\Coxeterc] has a complete list of equivalents).
The degrees are $d_1=q$, $d_2=1$, so $c=1$, $g=q$ and
$d_0=q$. There are no missing integers $\nu_i$ and, furthermore, $p_2=0$.
The fundamental domain is the lune, or digon, $(qq1)$.

For $q=1$ there is a single, equatorial
reflection plane, the fundamental domain being a hemisphere, (111) (a
spherical triangle with every angle equal to $\pi$). An alternative
notation for this domain is ${\bf A}_1$, [\Coxetera].
In this extreme case, $p_1$ is also zero and the expressions rapidly
collapse to
$$\ln\rho_2(1,1)=-\ze'_R(-1)-\ln\sqrt(2\pi),\quad \ze_2'(0,1\mid1,1)
=\ze'_R(-1)$$ and
$$\ze'(0,2\mid1,1)=\ze'_R(-1)+\ln\sqrt(2\pi).$$

Thus, on the hemisphere, from \zetaneutwo\ and \zetadirtwo,
\eqn\hsph{
\ze'_N(0)=2\ze'_R(-1)-\ln\sqrt(2\pi)-{1\over4},\quad\ze'_D(0)=
2\ze'_R(-1)+\ln\sqrt(2\pi)-{1\over4},
}
which agree with the results exhibited by Weisberger [\Weisbergerb]. Our
value of $\ze_N(0)=1/6$ does not agree with [\Weisbergerb].

The sum of the Neumann and Dirichlet expressions
should reduce to the full-sphere result derived by \eg Horta\c csu, Rothe
and Schroer [\Hortacsu] and later by Weisberger [\Weisbergera]. We find
$$\ze'_{{\rm S}^2}(0)=4\ze'_R(-1)-{1\over2}\approx-1.161684575$$
agreeing with these earlier calculations.
There are many discussions on spheres bounded equatorially by spheres.

We give the explicit formulae for the next value of $q$, $q=2$,
corresponding to a quartersphere,
\eqn\qsph{
\ze'_N(0)=\ze'_R(-1)-\ln\sqrt(2\pi)-{1\over8},\quad\ze'_D(0)=
\ze'_R(-1)+\ln\sqrt(2\pi)-{1\over8}.
}
Adding these expressions gives half the full-sphere value.

The results for higher values of $q$ are shown in {\it Fig}.1, where we
plot the effective action $W=-\ze'(0)/2$.
It is shown in the appendix that
\eqn\diff{
\ze'_N(0)-\ze'_D(0)=-\ln(2\pi)
}
for all $[q]$, as born out by the numbers.

For completeness we record the values of $\ze(0)$ obtained from \confanom,
relevant for the conformal anomaly,
\eqn\cnfantwo{
\ze_{{}_N}(0)=\ze_{{}_D}(0)={1\over12q}(1+q^2).
}

We turn now to the extended dihedral group, $[q,2]$, of order
$4q$, obtained from $[q]$ by adding a perpendicular reflection. It is the
complete symmetry group of the dihedron.
(In [\Polya,\Meyer] this group is $\ssD_{qi}$ ($q$ even) and
$\ssD_q[\ssD_{2q}$ ($q$ odd). The Sch\"onflies equivalent is $D_{qd}$.)

If $q$ is odd, $c=1$, $d_1=q$, $d_2=2$, $d_0=q+1$ and $g=2q$,
while, if $q$ is even, $c=2$, $e_1=q/2$ and $e_2=1$.
For odd $q$, the missing integers are $1,3,\ldots,q-2$. There are no
missing integers if $q$ is even.

The fundamental domain is the spherical triangle $(22q)$.
When $q=1$ this domain is the quartersphere lune and the results coincide
with \qsph. The group isomorphisms
are $ [1,2]\cong[2]\cong[1]\ttimes[1]$ (or $\ssC_2[\ssD_2\cong
\ssD_1[\ssD_2$ since $\ssC_2\cong\ssD_1)$.

Generally one has $[q,2]\cong[q]\times[1]$, in particular,
$[2,2]\cong[1]\ttimes[1]\ttimes[1]$ which corresponds to three perpendicular
reflections with the eighthsphere, $(222)$, as fundamental domain.

The hemisphere, quartersphere and eighthsphere are the intersections
of S$^2$ with $(\oR^+\ttimes \oR^2)$, $(\oR^+\ttimes\oR^+\ttimes \oR)$ and
$(\oR^+\ttimes\oR^+\ttimes\oR^+)$, respectively. The positive real axis,
$\oR^+$, is the positive root
space of the SU(2) algebra, ${\bf A}_1$ (\cf [\Berard]).
{\it Fig}.2 displays values of $W$ for bigger orders.

The rotation part of $[q,2]$ is the complete symmetry group of the
regular $q$-gon, $\{q\}$, and is the dihedral group in
its guise as a group of rotations. Coxeter denotes it by $[q,2]^+$ and
Polya and Meyer by $\ssb D_q$. As stated, its structure is $[q,2]^+
\cong\gD_q$.
When $q$ is odd there is the curious isomorphism $[2q]\cong[2,q]$.

It is only a matter of substitution to work out the the values of
\zetaneutwo\ and \zetadirtwo\ for the other reflection groups which are the
complete symmetry groups of the spherical tessellations $\{3,3\}$, $\{3,4\}$
and $\{3,5\}$. We find
$-\ze'(0)/2$ for (Dirichlet, Neumann)--conditions to be
$(0.45603, -0.34216)$ for $T_d=[3,3]$, $(0.2508, 0.001915)$ for
$O_h=[3,4]$ and $(-0.10538, 0.45014)$ for $I_h=[3,5]$.

The fundamental domain of $[p,q]$ is the spherical triangle $(pqr)$. The
rotational part of $[p,q]$, \ie $[p,q]^+$, is often denoted by $(p,q,r)$.
\sect{\bf 7. The Ces\`aro-Fedorov formula}.

\noin It is interesting to check the formula \confanom\ by remembering that
$\ze(0)$ is a local object
related to the constant term in the short-time expansion of the heat-kernel.
The general formula for a two-dimensional domain, $\cal M$, with boundary
$\pa{\cal M}=\bigcup\,\pa{\cal M}_i$ is
\eqn\beeone{
\ze(0)={1\over24\pi}\int_{\cal M} R\,dA+{1\over12}\sum_i\int_{\pa{\cal M}_i}
\ka(l)\,dl+{1\over24\pi}\sum_{\al}{\pi^2-\al^2\over\al}
}
where the $\al$ sum runs over all inward facing angles at the corners
of $\pa\cal M$.

In the present case $R=2$ and the extrinsic curvature, $\ka$, vanishes since
the boundaries of the fundamental domains are geodesic. Therefore
\eqn\zetzero{
\ze(0)={1\over24}\big({2\over g}+p+q+r-1\big)
}
where we have used the standard formula for the area of a spherical
triangle. This agrees with \confanom\ if the formula
\eqn\cesaro{
2d_0(d_0-1)=g(p+q+r-3)
}
is taken into account. In fact our derivation can be thought of as an
analytical proof of this relation which is a slight generalisation of
equation ${\bf 4\!\cdot\!51}$ in [\Coxetera]. (Coxeter has $r=2$ and $g=2N_1$.)

Coxeter indicates a purely geometric proof and points out that \cesaro\ is
equivalent to a formula discovered numerologically by Ces\`aro
[\Cesaro] and is a special case of an earlier result of Fedorov [see \Fedorov].
\cesaro\ is virtually identical to the equation on p177 of [\Cesaro] with the
correspondances $X=d_0,\,n=r,\,p=p$ and $q=q$.
An  extension to higher dimensions is possible using the generalisation
of \beeone\ that includes the results of Fedosov on polyhedral domains,
[\Fedosov].
\sect{\bf 7. Scaling and limits}.

\noin The results given so far are for a unit sphere. For radius
$R$, simple scaling gives the relation
\eqn\scale{
\ze'(0;R)=\ze'(0)+2\ln R\,\ze(0)
}
where $\ze'(0)=\ze'(0;1)$ and $\ze(0)=\ze(0;1)=\ze(0;R)$.

The effective action should incorporate an arbitrary scaling length, $L$,
by
$$
W_L=-{1\over2}\ze'(0;R)+\ln L\,\ze(0)=-{1\over2}\ze'(0)+\ln(L/R)\,\ze(0).
$$
The figures show just the first term.

Consider the dihedral case $[q]$ and let $q$ and $R$ tend to infinity in
such a way that the equatorial width of the fundamental domain $(qq1)$
remains fixed at $\be\equiv\pi R/q$. From \scale\ and \cnfantwo, whence
$\ze(0)\rightarrow q/12$, we have
\eqn\limit{
\ze'(0;R)\rightarrow\lim_{q\rightarrow\infty}\ze'(0)+{q\over6}\ln
\big({\be q\over\pi}\big).
}

The area of $(qq1)$ is $A_q=2\be^2q/\pi$ and requiring the
density, $\ze'(0;R)/A_q$, to remain finite as $q\rightarrow
\infty$ entails the leading behaviour
\eqn\leada{
\ze'(0)\rightarrow-{q\over6}\ln q +O(q).
}
Numerically we find
\eqn\leadb{
\ze'(0)\rightarrow-{q\over6}\ln q +0.497509 q\approx-{q\over6}\ln(q/19.79)
}
so that the density becomes
\eqn\effden{
{\ze'(0;R)\over A_q}\rightarrow{\pi\over12\be^2}\ln(6.299\be).
}

Geometrically, it might be imagined that in the limit $R=\infty$, since the
sphere becomes flat, the rescaled lune, $(\infty\infty1)$, would be
an infinite strip of width $\be$. Defining the strip coordinates
$x=R\phi$ and $y=R(\pi/2-\th)$, the spherical Laplacian does become the
usual Cartesian one as $R\rightarrow\infty$. However, the influence of the
infinitely sharp corners at the poles persists, even though they are
infinitely distant, producing an anomaly density of $\pi/12\be^2$. On the
rectangular strip, infinite or not, the integrated anomaly equals $1/2$ and
so the density vanishes in the infinite case.
\sect{\bf 8. The three-sphere.}

\noin
The expressions for the three-sphere are \zetaneuthree\ and \zetadirthree.
Then we require,
$$\ze_3(s,a\mid{\bf d})=\sum_{\bf m}{1\over(a+m_1d_1+m_2d_2+m_3d_3)^s}.$$
We will not attempt to extract a single \zf\ as we did previously but
will just reduce the sum to a finite one over Hurwitz \zfs\ in a not very
symmetrical nor economic fashion.

The residue classes
$$m_2=d_1n_2+p_1,\quad m_1=d_2n_1+p_2$$
are introduced so that the denominator function reads
$(a+d_1d_2(n_1+n_2)+p_2d_1+p_1d_2+m_3d_3)$. The sums over $n_1$ and $n_2$ can
be transformed by defining $n=n_1+n_2$ and doing the sum over
$n_1-n_2$ to yield the intermediate form
$$\ze_3(s,a\mid{\bf d})=\sum_{p_1,p_2}\sum_{n,m_3=0}^\infty{1+n\over
(a+d_1d_2n+p_2d_1+p_1d_2+m_3d_3)^s}.$$
The further residue classes
$$m_3=d_1d_2n_3+p_3,\quad n=d_3n_4+p_4$$
are introduced and the sum and difference defined by
$$n_+=n_4+n_3,\quad n_-=n_4-n_3.$$
The denominator is independent of $n_-$ while the numerator equals
$1+d_3(n_++n_-)/2+p_4$. Since the range of $n_-$ is symmetrical about zero
(from $-n_+$ to $n_+$ in steps of 2) the $n_-$ term gives nothing and there
is a factor of $(1+n_+)$ multiplying the rest. The sum may therefore be written
$$\ze_3(s,a\mid{\bf d})=\sum_{{\bf p},n}{(1+n)(1+d_3n/2+p_4)
\over(f+g n)^s}$$where ${\bf p}=(p_1,p_2,p_3,p_4)$,
$f=a+d_1d_2p_4+p_2d_1+p_1d_2+p_3d_3$, $g=d_1d_2d_3$ and we have set
$n=n_+$ for notational simplicity.

The numerator is reorganised to
$$(1+n)(1+d_3n/2+p_4)={d_3\over2g^2}\big(F+G(f+gn)+(f+gn)^2\big)$$ where
$$F=(A-g+d_1d_2p_4)(A-d_1d_2p_4-2d_1d_2),\quad G=g+2d_1d_2-2A$$
with $A$ being the combination $A=a+d_1p_2+d_2p_1+d_3p_3.$

Thus, finally, we arrive at a finite sum of Hurwitz \zfs,
\eqn\zetathree{\ze_3(s,a\mid{\bf d})={d_3\over2g^2}\sum_{\bf p}
\left[{F\over g^s}\ze_R\big(s,{f\over g}\big)+
{G\over g^{s-1}}\ze_R\big(s-1,{f\over g}\big)+{1\over g^{s-2}}
\ze_R\big(s-2,{f\over g}\big)\right]
}
which constitutes a possible, but inefficient, continuation of the
Barnes \zf.
\sect{\bf 9. The honeycomb groups}.

\noin The three-dimensional analogues of the polyhedral tessellations,
$\{p,q\}$, of the two-sphere are the spherical honeycombs $\{p,q,r\}$,
[\Coxetera,\Coxeterb,\Coxeterd]. The reflection groups $[p,q,r]$ are their
complete symmetry groups, the fundamental domains being subspaces of the
honeycomb cells.  A numerical calculation using \zetathree\ and
\zetadirthree\ produces the following typical results for the Dirichlet
effective actions. For $[3,3,3]$, $W\approx 44.4$ and for $[3,3,4]$,
$W\approx -427.25$.

\sect{\bf 10. Conclusion}.

\noin The results of this paper are strictly technical. We have achieved
our aim of presenting calculable formulae for the functional determinants
of minimally coupled scalar fields on the fundamental domains of finite
reflection groups. The problem has devolved upon an evaluation of the
derivative of the Barnes \zf.

We could also extend our previous results on the vacuum energies
[\Chang] to minimal coupling using the expressions for $\ze(-n)$,
\zetaminusn, and $\ze'(-n)$. This straightforward exercise will not be
done here.

The Ces\`aro-Fedorov formula for the number of symmetry planes of
a regular solid proved in section 6 is one of a number of similar relations in
higher dimensions derivable from expressions for the coefficients in the
short-time expansion of the heat-kernel. The details will be presented
elsewhere.

The conformal transformations taking a fundamental domain into the upper
half--plane are known and so the results here described should also be
obtainable using standard conformal techniques. This will be recounted
at another time.
\vfill\eject
\sect{\bf Appendix}.

\noin In this appendix we first work out an expression for the derivative of
the \zf\ \genzet\ at
negative integers, $\ze'(-n)$. For brevity we do not display the dependence of
the Barnes \zf\ on the ${\bf d}$.

Differentiation of \binsum\ first of all leads to
$$
\ze'(-n)=\sum_{r=0}^n(-\al^2)^r\left(n\atop r\right)\left[2\ze_d'(2r-2n,a)-
\ze_d(2r-2n,a)\sum_{k=n-r+1}^n{1\over k}\right]+$$
$$\phantom{{\hbox{*********}}}(-1)^n\sum_{r=n+1}^{n+u}\al^{2r}{n!(r-n-1)!
\over r!}\left(R_{2r-2n}+
{1\over2}N_{2r-2n}\sum_{k=n+1}^{r-n-1}{1\over k}\right)+$$
\eqn\zdashm{
(-1)^n\sum_{r=u+1+n}^\infty\al^{2r}{n!(r-n-1)!\over r!}\ze_d(2r-2n,a).
}

We substitute the integral form of the Barnes \zf\ into the last term and find
it as the $s\rightarrow-2n$ limit of $2^n(-1)^nn!$ times

\eqn\intint{
2\int_0^\infty\exp(-a\tau)\left(\cosh\al\tau-\sum_{r=0}^{n+u}{(\al\tau)^{2r}
\over(2r)!}\right){\tau^{s-1}d\tau\over\prod_i\big(1-\exp(-d_i\tau)\big)}
}
which equals
\eqn\intintint{
\Ga(s)\big(\ze_d(s,a-\al)+\ze_d(s,a+\al)\big)-
2\sum_{r=0}^{u+n}{\al^{2r}\over(2r)!}\Ga(s+2r)\ze_d(s+2r,a).
}
The pole cancellation gives the condition

$$
\ze_d(-2n,a-\al)+\ze_d(-2n,a+\al)-2\ze_d(-2n,a)=\phantom{\hbox{ZZZZZZZZZZZZZZ}}
$$

\eqn\polen{=2\sum_{r=1}^n\al^{2r} \comb{2n}{2r} \ze_d(2r-2n)+
2\sum_{r=n+1}^{n+u}\al^{2r}{(2r-2n-1)!\over(2r)!}N_{2r-2n}.
}

Extracting the finite remainder yields, after using \polen,

$${1\over(2n)!}\bigg(\ze'_d(-2n,a-\al)+\ze'_d(-2n,a+\al)-2\ze'_d(-2n,a)\bigg)+$$
$${2\over(2n)!}\sum_{r=1}^n\al^{2r}\comb{2n}{2r}\bigg[\big(\psi(1+2n-2r)-
\psi(1+2n)\big)\ze_d(2r-2n,a)-\ze_d'(2r-2n,a)\bigg]$$
\eqn\finrem{-2\sum_{r=n+1}^{n+u}\al^{2r}{(2r-2n-1)!\over(2r)!}
\bigg[\big(\psi(2r-2n)+\psi(1+2n)\big)N_{2r-2n}+
R_{2r-2n} \bigg].
}

Multiplied by $2^n(-1)^nn!$, \finrem\ must be substituted into \zdashm\ to
yield a calculable formula for $\ze'(-n)$. Doing so reveals that the
remainder terms $R_{2r-2n}$ cancel but, apart from this, there are no
other simplifications apparent and we leave the analysis at this point.

We next derive the result \diff\ starting from \funcda\ and
\funcdb\ whence
\eqn\diffa{
e^{-\ze'_N(0)+\ze'_D(0)}={\Ga_2(d_0)\Ga_2(d_0+1)\over\Ga_2(1)}.
}
It is necessary to use some properties of the multiple $\Ga$--function.

{}From \modform\ and \multgam\ it is obvious that
\eqn\limgam{
\lim_{a\rightarrow0}\Ga_r(a)={1\over a}.
}
The other properties we need follow from the important recursion formula
satisfied by the Barnes \zf,
\eqn\recn{
\ze_r(s,a+d_i\mid{\bf d})-\ze_r(s,a\mid{\bf d})=-\ze_{r-1}(s,a\mid{\bf d}'),
}
where ${\bf d}'$ stands for the set of degrees ${\bf d}$ with the $d_i$
element omitted.

If this equation is differentiated, it quickly results that, [\Barnes],
\eqn\gamrat{
{\Ga_r(a)\over\Ga_r(a+d_i)}={\Ga_{r-1}(a)\over\rho_{r-1}({\bf d}')}.
}
Setting $a$ equal to zero in \gamrat\ and using \limgam\ it follows that
\eqn\spgam{
\Ga_r(d_i)=\rho_{r-1}({\bf d}').
}

For the group $[q]$, we recall that the degrees are ${\bf d}=(q,1)$.
Then, choosing $d_i=d_1=q$ and setting $a=1$, we have from \gamrat
$${\Ga_2(1)\over\Ga_2(1+q)}={\Ga_1(1)\over\rho_1(1)}$$ which is clearly
independent of $q$ since the quantities on the right-hand side are calculated
for the degree ${\bf d}'=(1)$. Further, from \spgam, it is likewise
clear that $\Ga_2(q)$ is independent of $q$.
Therefore the quantity in \diffa,
$${\Ga_2(q)\Ga_2(q+1)\over\Ga_2(1)}={\rho^2_1(1)\over\Ga_1(1)},$$
is independent of $q$. The actual value is $2\pi$, agreeing
with the particular cases \hsph\ and \qsph.

Incidentally, from the general formulae \nzf, \dzf\ and \recn\ it easily
follows that the $[q]$ conformal \zfs\ are related by
\eqn\rec{
\ze^{(C)}_{{}_N}(s)-\ze^{(C)}_{{}_D}(s)=
\ze_1(2s,1/2\mid1)=\ze_R(2s,1/2)
}
so that, in particular,
$$
{\ze^{(C)}_{{}_N}}'(0)-{\ze^{(C)}_{{}_D}}'(0)=
-\ln(2)$$ for all $[q]$.

  \vfill\eject\immediate\closeout\reffile
  \centerline{{\bf References}}\bigskip\frenchspacing%
  \input refs.tmp\vfill\eject\nonfrenchspacing

\end